\documentclass[12pt,a4paper,twoside,prflogo,Research,Versita]{versita}
\def\apj{ApJ\,}
\def\aap{A\&A\,}
\def\mnras{MNRAS\,}
\def\aj{AJ}

\year{2007}
\issue{5(1)}

\def\sun{\hbox{$\odot$}}
\def\2F1{~_2F_1}

\year{2007}
\issue{Physics  ..}
\rightmark{ L. Zaninetti and M. Ferraro }
\leftmark{  L. Zaninetti and M. Ferraro}
\title{
On the Truncated  Pareto  Distribution
with applications 
}
\author{ L. Zaninetti \inst{1}\email{E-mail: zaninetti@ph.unito.it} ,
         M. Ferraro   \inst{2}\email{E-mail:   ferraro@ph.unito.it}
            }
\institute{
 Dipartimento  di Fisica Generale,\\
Universit\`a degli Studi di Torino    \\
via P.Giuria 1,  10125 Torino,Italy    \\
and                                   \\
 Dipartimento  di Fisica Sperimentale,\\
Universit\`a degli Studi di Torino    \\
via P.Giuria 1,  10125 Torino,Italy 
}
\received{  }
\revised {  }
\abstract{
The  Pareto probability distribution is widely applied 
in different fields such us finance, physics
, hydrology , geology  and astronomy. 
This note deals with an application  of the Pareto distribution to astrophysics
 and more precisely to the statistical analysis of 
mass of stars and of diameters of asteroids. In particular 
a comparison between the usual Pareto distribution and 
its truncated version is presented.
Finally a  possible physical mechanism that produces 
Pareto tails for the
distribution of the masses of stars is suggested.
}
\keyword{ statistical distributions ,stars, asteroids }
\PACS {
95.75.-z Observation and data reduction techniques; computer modelling and
simulation ;
96.30.Ys 	Asteroids, meteoroids
97.10.Xq        Stars: 	Luminosity and mass functions
}

\begin{document}
\firstpage{1}
\maketitle \setcounter{page}{1}%

\section{Introduction}
The Pareto distribution  \cite{Pareto,evans}.
is a simple model
 for nonnegative data with a power law probability tail. 
In many practical applications, it is natural to consider an upper bound that 
truncates the tail  \cite{Cohen1988,Devoto1998,Aban2006};
the truncated Pareto distribution has a wide range of applications 
in several fields in  
data analysis~\cite{Aban2006} \cite{Rehfeldt1992}.

Power law distributions are often found in astrophysics: 
for instance 
 in the range
$1 \mathcal {M}_{\sun}~<~\mathcal {M} <  10 \mathcal {M}_{\sun}$,
the mass of the stars (MAIN SEQUENCE V), when expressed
in terms of the solar mass $\mathcal {M}_{\sun}$,
 scales as
$\psi({\mathcal M})\propto  {\mathcal{M}}^{-\alpha}$
with  $\alpha$= 2.35,
see~\cite{Salpeter1955} or $\alpha$= 2.3
as  suggested by 
a recent evaluation , see~\cite{Kroupa2001}.
 Other examples are the intensity  of  non thermal
emission  from supernova remnant and extra-galactic radio-sources 
that scales as 
$\nu^{-\alpha}$,  with numerical 
values of $\alpha$  ranging between 0.5 and 1, 
the observed differential spectrum of cosmic rays
proportional to  $E^{-2.75}$ in the interval $10^{10}~eV-5.0~10^{15}~eV$
\cite{lang,Schlickeiser},
the gamma ray bursts luminosity   function 
that scales as
$L^{-2}$, 
\cite{Rossi2005,Bloom2001}.
Of course the Pareto distribution  is not the only one to exhibit  
power law tail, this behaviour being common to different 
distributions (e.g. the lognormal distribution); 
however Pareto distributions are
specially attractive for their simple analytical form. 

In this paper we present in Section~\ref{preliminar}
a comparison between the Pareto 
and the truncated Pareto distributions.
In Section~\ref{applications}  the theoretical results are 
 applied to 
distribution of astrophysical data, namely the mass of stars and the 
radius of asteroids.  
A physical mechanism  that produces a Pareto type
distribution for the masses is presented in Section~\ref{tails}

\section{Preliminaries}
\label{preliminar}
Let $X$ be a random variable taking values $x$ in the interval 
$[a, \infty]$, $a>0$.
The  probability density function (in the following pdf)
named Pareto  
is defined by  \cite{evans}
\begin {equation}
f(x;a,c) = {c a^c}{x^{-(c+1)}} \quad ,
\label{pareto}
\end {equation}
$ c~>0$,
and the Pareto distribution functions   is
$F(x:a,c)=1-a^cx^{-c}$

An upper truncated Pareto random variable is defined in the interval 
$[a,b]$ and the corresponding pdf is 
\begin {equation}
f_T(x;a,b,c ) = \frac{ca^cx^{-(c+1)}}{1-\left (\frac{a}{b}\right)^c}
\quad ,  \label{eq:pdf}
\end {equation}
\cite{Aban2006} and 
the truncated Pareto distribution function is  

\begin{equation}
F_T(x;a,b,c) =
\frac {1 -(\frac {a}{x})^c} {1-(\frac{a}{b})^c}
\quad .
\end{equation}

Momenta of the truncated distributions exist for all $c>0$. For instance,
the 
mean of   $f_T(x; a,b,c)$ is, for  $c\neq 1$ and $c=1$, respectively,
\begin{equation}
\label{eq:avtr}
<x> =
\frac {ca}{c-1} 
\frac {1 -(\frac{a}{b})^{c-1}}{1-(\frac{a}{b})^c}, \quad 
<x> =\frac{ca^c}{1-\left (\frac{a}{b}\right )^c}\ln\frac{b}{a}
\end{equation}

Similarly, if $c\neq 2$, the variance is given by
\begin{equation}
\label{eq:vart}
\sigma^2 = \frac{ca^2}{(c-2)}\frac{1-\left (\frac{a}{b} \right)^{c-2}}
{1-\left (\frac{a}{b} \right )^c} -<x>^2,
\end {equation}

whereas for $c=2$
\begin{equation}
\label{eq:svart}
\frac{ca^c}{1-\left (\frac{a}{b}\right )^c}\ln\frac{b}{a}-<x>^2.
\end{equation} 

In general the  $n-th$ central moment is  
\begin{eqnarray}
\int_a^b(x-<x>)^n f_T(x)dx=  \nonumber \\
 \left( -{\it <x>} \right) ^{n}{a}^{-c}
{\2F1(-c,-n;\,1-c;\,{\frac {a}{{\it <x>}}})} \left(  \left( {a}^{c}
 \right) ^{-1}- \left( {b}^{c} \right) ^{-1} \right) ^{-1}
\nonumber \\
- \left( -{\it <x>} \right) ^{n}{b}^{-c}
{\2F1(-c,-n;\,1-c;\,{\frac {b}{{\it <x>}}})} \left(  \left( {a}^{c}
 \right) ^{-1}- \left( {b}^{c} \right) ^{-1} \right) ^{-1}
\end{eqnarray}
where ${\2F1(a,b;\,c;\,z)}$ is a regularized hypergeometric function,
see~\cite{Abramowitz1965,Seggern1992,Thompson1997}.
An analogous formula
 based
on some of the properties of the incomplete beta function, 
see~\cite{Grad2000}  and \cite{Prud1986} ,
can be  found in~\cite{Ali2006}.

Parameters of the truncated Pareto pdf 
from empirical data can be obtained via  
the maximum likelihood method; explicit formulas 
for maximum likelihood estimators  (MLE) are given in \cite{Cohen1988}, 
and for  the more general case  in 
 \cite{Aban2006}, whose results we report here for completeness. 

Consider a random sample  ${\mathcal X}=x_1, x_2 , \dots , x_n$ and let 
$x_{(1)} \geq x_{(2)} \geq \dots \geq x_{(n)}$ denote 
their order statistics so that 
$x_{(1)}=\max(x_1, x_2, \dots, x_n)$, $x_{(n)}=\min(x_1, x_2, \dots, x_n)$.

The MLE of the parameters $a$ and $b$ 
are
\begin{equation}
{\tilde a}=x_{(n)}, \qquad {\tilde b}=x_{(1)},
\label{eq:firstpar}
\end{equation}  

respectively, and $\tilde c$ is the solution of the equation

\begin {equation}
\label{equationmle}
\frac {n}{{\tilde c}} +
\frac {n  (\frac {x_{(n)}}{x_{(1)}})^{\tilde c} 
\ln (\frac{x_{(n)}}{x_{(1)}})}{ 1-(\frac{x_{(n)}}{x_{(1)}})^{\tilde c }}
- \sum_{i=1}^n  [\ln x_i-\ln x_{(n)}]
= 0,
\end {equation}
\cite{Aban2006}.

There exists a 
simple test to see whether a Pareto model is appropriate \cite{Aban2006}:
the null hypothesis $H_0:\nu=\infty$ is rejected if and only if 
$x_{(1)} < [nC/(-\ln q)]^{1/c}$, $0<q<1$, where $C=a^c$.
The approximate $p$-value of this test is given 
by $p=\exp\left \{-nCx^{-c}_{(1)}\right \}$, and a small value of $p$ 
indicates that the 
Pareto model is not a good fit; of course this is not enough  
{\it per se} to demonstrate the goodness of a truncated Pareto distribution.

Given a set of data is  often difficult to decide if 
they agree more closely with $f$ or $f_T$, in that, in the interval 
$[a,b]$, they differ only or a multiplicative factor 
$1-(a/b)^c$, that if the interval $[a, b]$
is not too small approaches $1$ even for relatively small 
values of $c$.
For this reason,  rather than $f$ and $f_T$,  
the distributions $P(X>x)$ and $P(X>x)$ are used,
often called survival functions,
that are given respectively by

\begin{equation}
\label{eq:surv}
P(X>x)=S(x)=1-F(x;a,c)=a^cx^{-c}
\end{equation}

and 
\begin{equation}
\label{eq:surt}
P_T(X>x)=S_T(x)=1-F_T(x;a,b,c)=
\frac{ca^c \left(x^{-c}-b^{-c}\right)}{1-\left(\frac{a}{b}\right)^c}.
\end{equation}

\begin{figure}
{
\includegraphics[width=12cm]{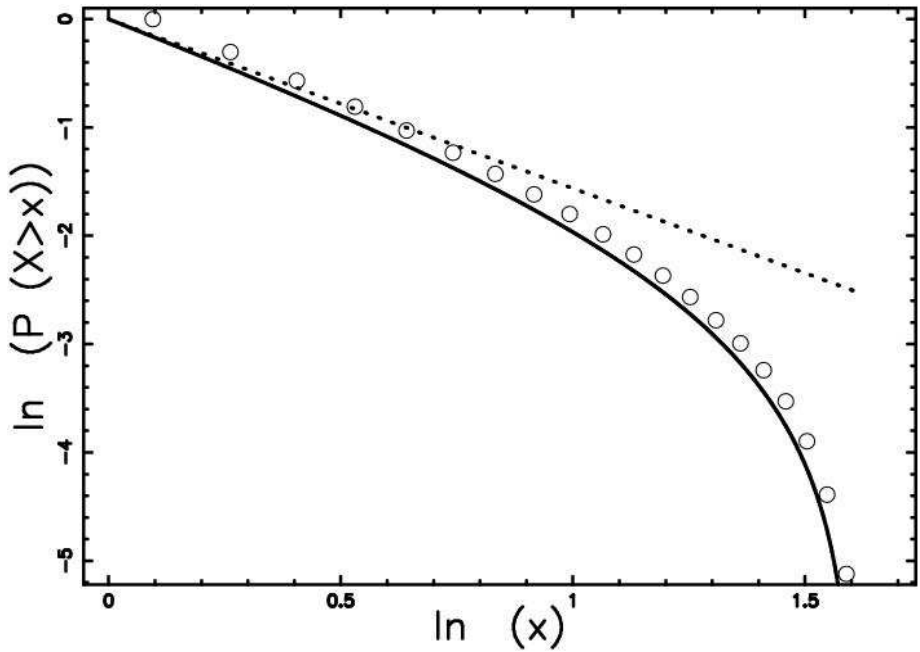}
}
\caption{
log--log plot  of the survival function:
10000 random data (empty circles), generated with Eq. (\ref{eq:simpar}), 
survival function  of the truncated Pareto distribution (full line) and
survival function  of the           Pareto distribution (dotted line).
}
\label{simulation}
\end{figure}

Probabilities $P$ and $P_T$ have qualitatively
different trends  that are better observed 
in a log-log plot. In this case $P$ is obviously represented by 
a straight line, whereas $P_T$ exhibits also a almost linear trend
with a sharp drop when $x$ tends to $b$.
To illustrate this point we have generated a set of 
$n=10000$ random points drawn 
from a truncated Pareto distribution, 
via the formula 
\begin{equation}
X  : a,b,c   \sim
a\left( 1 - R ({1-(\frac{a}{b})^c)}  \right) ^{-\frac{1}{c}}
\quad ,
\label{eq:simpar}
\end{equation}
where $ R$ is the unit rectangular variate, and we have fitted them 
with $S$ and $S_T$ respectively, 
see Figure~\ref{simulation}.

\section{Applications}
\label{applications}
\subsection{Mass of stars}

\label{secapplications} 

\begin{figure}
{
\includegraphics[width=12cm]{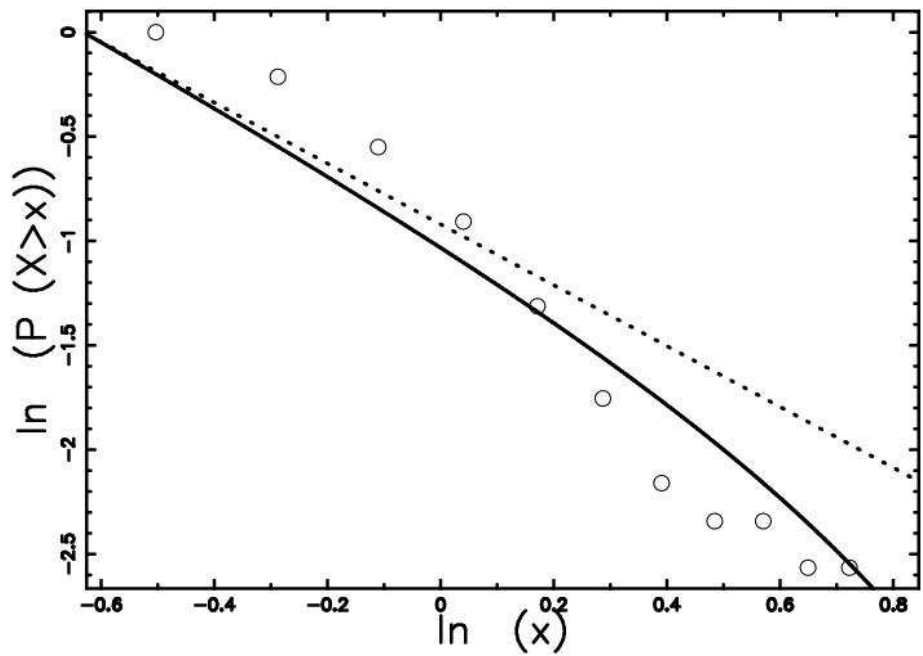}
}
\caption{
log--log plot  of the survival function 
of the   mass distribution of the stars:
data (empty circles),
survival function  of the truncated Pareto  pdf (full line) and
survival function  of the           Pareto  pdf (dotted line).
A  complete sample (MAIN SEQUENCE V) is considered with
parameters as in Table~\ref{stars_complete}.
}
\label{pareto_greater05}
\end{figure}

The sample of star's  masses has been 
obtained from the Hipparcos  data as a function of the absolute
magnitude and (B-V)~\cite{zaninetti05}. 
Results of the fitting with $P$ and $P_T$ are presented 
in Table~\ref{stars_complete} where $a,b,c$ and $n$, the number of sample elements, are reported and in 
 Figure~\ref{pareto_greater05} that shows
 the data with
 the fit.
\begin{table}[h]
\caption{\it Coefficients of  mass distribution of the stars
         in the first 10~pc,  of
         a  complete sample (MAIN SEQUENCE V).
         The parameter $c$ is derived through MLE and p=0.032.}
\label{stars_complete}
\begin{tabular}{ccccc}
\hline
a [$\mathcal {M}_{\sun}$]         &  b [$\mathcal {M}_{\sun}$]      &   c    &  n
& $P(X>x)$         \\
\hline
0.53       &  3.44   & 1.45   &   52        & Truncated~Pareto                  \\
\hline
0.53       &  $\infty $  &  1.77   &   52    & ~Pareto                  \\
\hline
\hline
\end{tabular}
\end{table}

In this  case  in the range
$3.44 \mathcal {M}_{\sun}~>~\mathcal {M} \geq  0.53  \mathcal {M}_{\sun}$,
see Table~\ref{stars_complete} , the coefficient $\alpha=c+1$
=2.45 is in agreement with modern estimates~\cite{Kroupa2001}.
In this case, the power of the Pareto test results to be $p=0.032$,
indicating that the Pareto distribution is not a good fit, 
as can also be seen from Figure~\ref{pareto_greater05}   .

Table~\ref{stars_chi2}  therefore  reports the $\chi^2$ of the
fit of the stars when  the Pareto and the truncated Pareto, respectively.

\begin{table}[h]
\caption{
 $\chi^2$ 
of  different distributions when the
 number of bins is 5 for the stars in the first 10 $pc$~.}
\label{stars_chi2}
\begin{tabular}{lcc}
\hline
Distribution  & $\chi^2$  \\
\hline
Pareto               &  7.1       \\
Truncated~Pareto     &  5.26      \\
\hline
\hline
\end{tabular}
\end{table}

\subsection{Distribution of asteroids size}

Suppose that not just masses of stars but also 
those of other astrophysical objects have a power law tail, 
then is not difficult to prove that also their  linear dimension, 
radii or diameters, 
must follows a power law.
We have tested this hypothesis by considering diameters 
of different families of asteroids, namely, Koronis , Eos and Themis.

\begin{figure}
{
\includegraphics[width=12cm]{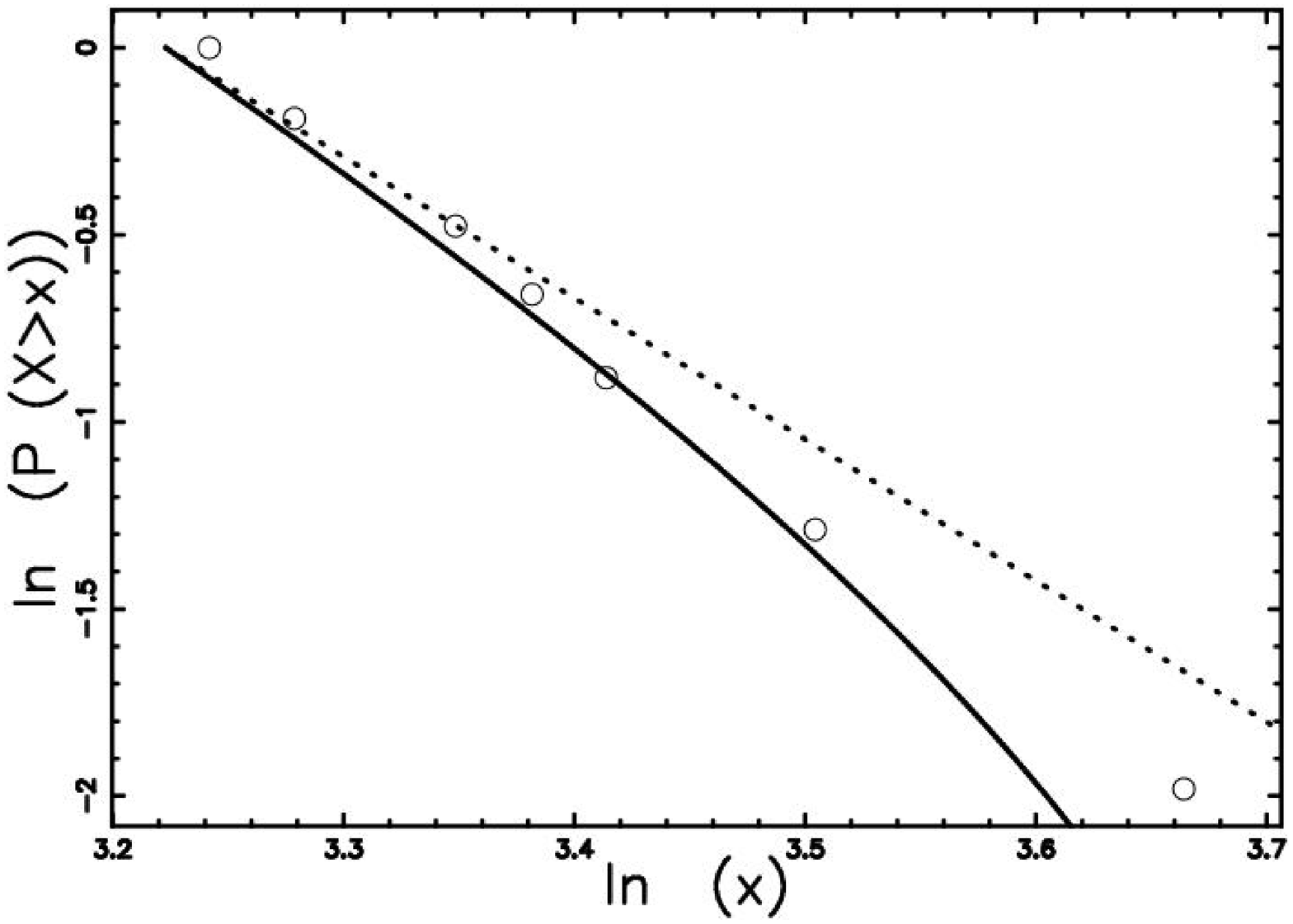}
}
\caption{
ln--ln plot  of the survival function 
of  diameter distribution of the Koronis  Family:
data (empty circles),
survival function  of the truncated Pareto  pdf (full line) and
survival function  of the           Pareto  pdf (dotted line).
A  complete sample is considered with
parameters as in Table~\ref{Koronis}.
}
\label{pareto_tronc_koronis_diam}
\end{figure}

\begin{figure}
{
\includegraphics[width=12cm]{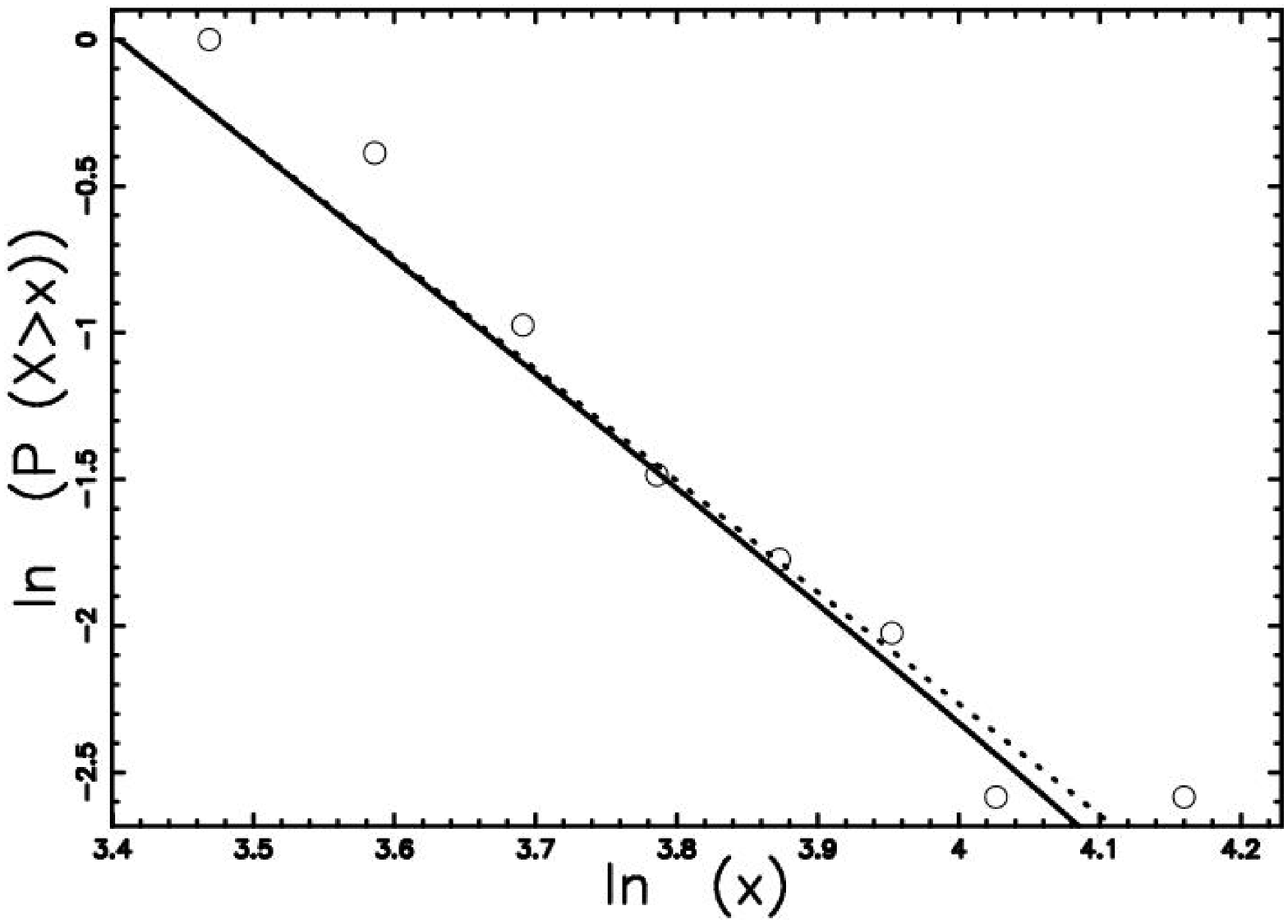}
}
\caption{
ln--ln plot  of the survival function 
of  diameter distribution of the Eos Family:
data (empty circles),
survival function  of the truncated Pareto  pdf (full line) and
survival function  of the           Pareto  pdf (dotted line).
A  complete sample is considered with
parameters as in Table~\ref{Eos}.
}
\label{pareto_tronc_eos_diam}
\end{figure}

\begin{figure}
{
\includegraphics[width=12cm]{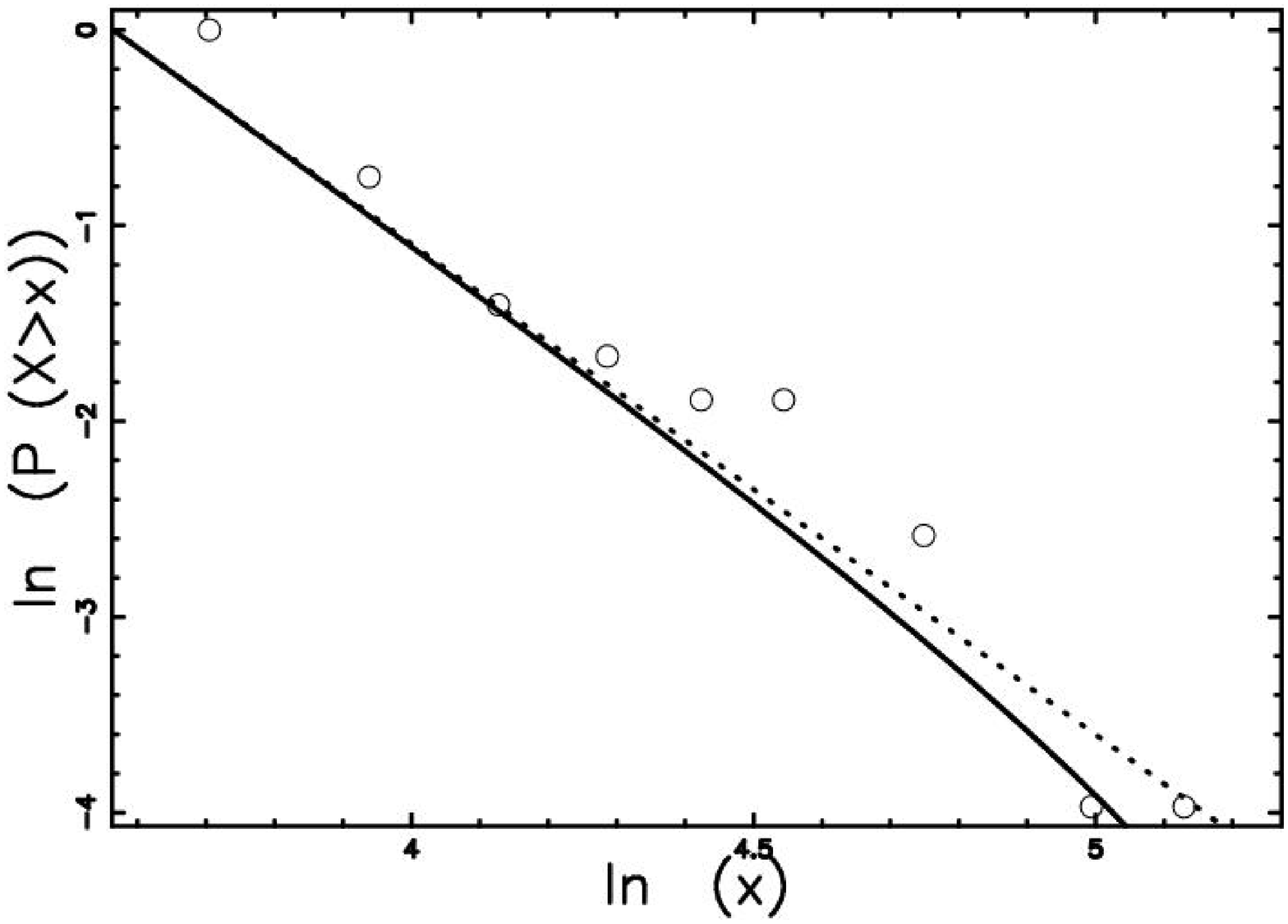}
}
\caption{
ln--ln plot  of the survival function 
of  diameter distribution of the Themis  Family:
data (empty circles),
survival function  of the truncated Pareto  pdf (full line) and
survival function  of the           Pareto  pdf (dotted line).
A  complete sample is considered with
parameters as in Table~\ref{Themis}.
}
\label{pareto_tronc_themis_diam}
\end{figure}

In the following The  sample parameter of the families are 
reported in Table~\ref{Koronis},Table~\ref{Eos} and Table~\ref{Themis} ,
 whereas 
Figure~\ref{pareto_tronc_koronis_diam},
Figure~\ref{pareto_tronc_eos_diam},
Figure~\ref{pareto_tronc_themis_diam} 
report the graphical display of data and the fitting distributions.

\begin{table}[h]
\caption{\it Coefficients of  diameter  distribution of the Koronis family .
         The parameter $c$ is derived through MLE and p=0.033 .}
\label{Koronis}
\begin{tabular}{ccccc}
\hline
a [km]         &  b [km]      & c       &  n  &  $P(X>x)$                \\
\hline
25.1           &  44.3        & 3.77    &  29 & truncated ~Pareto                    \\
\hline
25.1           & $\infty$     & 5.04    &  29 & Pareto                    \\
\hline
\hline
\hline
\end{tabular}
\end{table}

\begin{table}[h]
\caption{\it Coefficients of  diameter  distribution of the Eos family .
         The parameter $c$ is derived through MLE and p=0.681 .}
\label{Eos}
\begin{tabular}{ccccc}
\hline
a [km]         &  b [km]      & c       &  n  &  $P(X>x)$                \\
\hline
30.1           &  110        & 3.80    &  53 & truncated ~Pareto                    \\
\hline
30.1           & $\infty$    & 3.94    &  53 & Pareto                    \\
\hline
\hline
\hline
\end{tabular}
\end{table}

\begin{table}[h]
\caption{\it Coefficients of  diameter  distribution of the Themis family .
         The parameter $c$ is derived through MLE and p=0.67 .}
\label{Themis}
\begin{tabular}{ccccc}
\hline
a [km]         &  b [km]      & c       &  n  &  $P(X>x)$                \\
\hline
35.3           &  249         & 2.5     &  53  & truncated ~Pareto                    \\
\hline
35.3           & $\infty$     & 2.6     &  53 & Pareto                    \\
\hline
\hline
\hline
\end{tabular}
\end{table}

In case of the Koronis family $P_T$ fits the data better than 
$P$ and indeed $p=0.039$ is correspondingly small, whereas 
for the Eos family,
$P$ performs slightly better than $P_T$ (p=0.68), and the estimated 
of $c$ are very closed in both cases. Finally in the third case, the Themis 
family, the two distributions are the same, due to the fact that 
the ratio $a/b=0.14$ is small.

\section{Generating Pareto  tails}
\label{tails}
As a  simple example of how 
a distribution with power can be generated, consider   
the growth of a primeval nebula via accretion, 
that is the process by which nebulae 
``capture'' mass. 
 We start by considering  an uniform pdf  for the initial mass
of $N$ primeval nebulae, $m$, in a  range 
  $m_{min} <    m \leq  m_{max} $ .
At each interaction the $i$-th nebula has a probability $\lambda_i$ 
to increase its mass $m_i$
 that is given by
\begin{equation}
\label{eq:trans}
\lambda_i=(1-\exp(-ak m_i)),
\end{equation}
where $ak$ is a parameter of the simulation;
thus more ``massive'' nebulae are more likely 
to grow, via accretion. 
The quantity of which the primeval nebula 
can grow varies with time, to take into account that the total mass 
available is limited,
\begin{equation}
\delta m(t)=\delta m(0)\exp(-t/\tau)
\label{eq:dm}
\quad ,
\end{equation}
where $\delta m(0)$ represents the maximum 
mass of exchange and $\tau$ the scaling time of the phenomena.
The simulation proceeds as follows: 
a number $r$,  is randomly chosen in the interval $[0,1]$ 
for each nebula, and, 
if $r < \lambda_i$, the mass $m_i$ is increased by 
$\delta m(t)$, where $t$ denotes the iteration of the process.
The process proceed in parallel : at each 
temporal iteration all the primeval nebulae 
are considered.

\begin{figure}
{
\includegraphics[width=12cm]{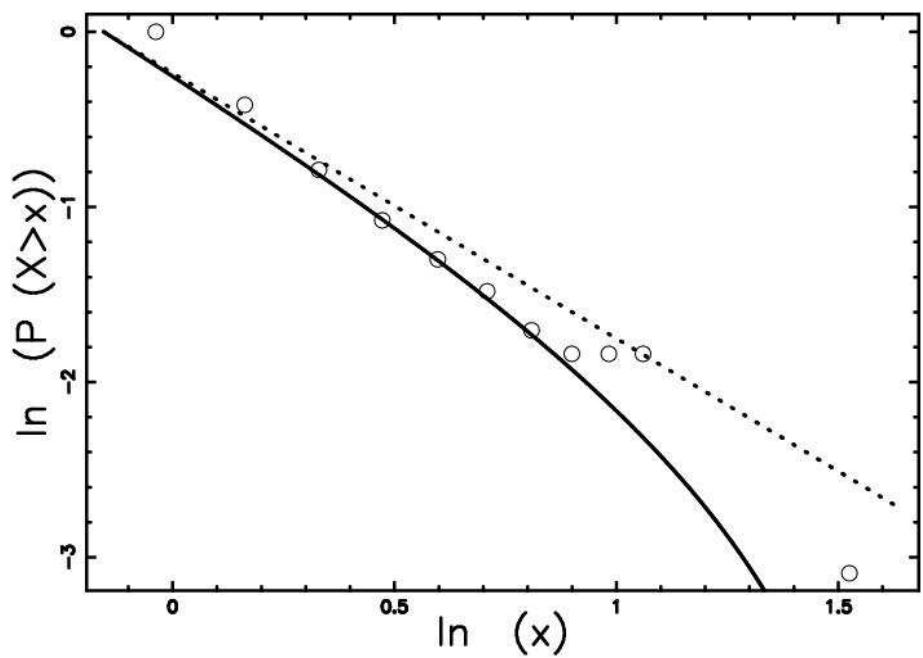}
}
\caption{
log--log plot  of the survival function 
of the   mass distribution for the primeval nebula when
$m \geq 0.5\mathcal {M}_{\sun} $ are considered.
The truncated Pareto parameters are  $c$=1.36  and p=0.0058~.
}
\label{masses_pareto}
\end{figure}

Results of the simulations  have been fitted with both
  Pareto survival distributions.
see Figure~\ref{masses_pareto}

Due to a photometric effect~\cite{zaninetti05} 
the sample of observed stars is complete only
for $m \geq 0.5\mathcal {M}_{\sun} $.
We  therefore have set the lower  boundary of the
masses to $0.5\mathcal {M}_{\sun} $, and the resulting 
subset  has been fitted with the 
Pareto and truncated Pareto survival distributions,
Figure~\ref{masses_pareto}. It should be noted that 
the results of the simulation give $c=1.36$, that is 
$\alpha=2.36$ in agreement with the experimental estimate.

\section{Conclusions}
Results of the analysis presented here  show that the truncated Pareto
distribution provides a good fit for the distribution 
and performs better than the usual Pareto distribution. 
When the asteroid diameters are considered the situation is not 
so clear in that it depends on the family one considers.
It is also clear the there can be cases, such as with the Themis family,
in which the ratio between the minimum and the maximum value 
of the sample is so small that there no real difference between the 
two distributions. 
Finally we have shown that Pareto distributions can result from 
a simple growth process, in which the increase of the state variable  
(here mass) depends on the values taken in the previous state; furthermore 
results of the simulations agree well with the experimental data. 

As remarked earlier 
distributions are not the only statistics with a power law tail; 
also in astrophysics  alternatives have been proposed for the statistics 
of asteroid diameters (e.g. \cite{Zaninetti1995}). 
However Pareto distributions are 
particularly simple; 
for instance note that they have just a free parameter $c$, 
the others $a$ and $b$, being determined by 
the minimum and maximum values of the sample, respectively.  


\end{document}